\documentclass[pdflatex,sn-mathphys-num]{sn-jnl}

\usepackage{graphicx}%
\usepackage{multirow}%
\usepackage{amsmath,amssymb,amsfonts}%
\usepackage{amsthm}%
\usepackage{mathrsfs}%
\usepackage[title]{appendix}%
\usepackage{textcomp}%
\usepackage{manyfoot}%
\usepackage{booktabs}%
\usepackage{algorithm}%
\usepackage{algorithmicx}%
\usepackage{algpseudocode}%
\usepackage{listings}%

\begin{document}

\title[Article Title]{Model-based deep reinforcement learning for accelerated learning from flow simulations}


\author*[1]{\fnm{Andre} \sur{Weiner}}\email{andre.weiner@tu-dresden.de}
\equalcont{These authors contributed equally to this work.}

\author[1]{\fnm{Janis} \sur{Geise}}\email{janis.geise@tu-dresden.de}
\equalcont{These authors contributed equally to this work.}

\affil*[1]{\orgdiv{Institute of Fluid Mechanics}, \orgname{TU Dresden}, \orgaddress{\street{George-B\"{a}hr-Str. 3c }, \city{Dresden}, \postcode{01069}, \state{Saxony}, \country{Germany}}}


\abstract{
In recent years, deep reinforcement learning has emerged as a technique to solve closed-loop flow control problems.
Employing simulation-based environments in reinforcement learning enables a priori end-to-end optimization of the control system, provides a virtual testbed for safety-critical control applications, and allows to gain a deep understanding of the control mechanisms.
While reinforcement learning has been applied successfully in a number of rather simple flow control benchmarks, a major bottleneck toward real-world applications is the high computational cost and turnaround time of flow simulations.
In this contribution, we demonstrate the benefits of model-based reinforcement learning for flow control applications.
Specifically, we optimize the policy by alternating between trajectories sampled from flow simulations and trajectories sampled from an ensemble of environment models.
The model-based learning reduces the overall training time by up to $85\%$ for the fluidic pinball test case.
Even larger savings are expected for more demanding flow simulations.

}




\maketitle

\section{Introduction}\label{sec:intro}
According to the 2022 IPCC report, smart control technologies are a key enabler in mitigating the impact of global warming and climate change \cite{ipcc2022}.
Control tasks in which the manipulation of fluid flows can significantly reduce the carbon dioxide footprint are ubiquitous in society and industries, e.g., reducing aerodynamic forces acting on vehicles \cite{seifert2015}, maximizing the coefficient of performance of heat pumps \cite{rohrer2023}, or adjusting the operating conditions of process engineering systems in response to the availability of renewable energy resources \cite{esche2020}, to name a few.
To achieve optimal control in the aforementioned example applications, it is necessary that the controller can respond to changes in the ambient conditions.
However, the design and implementation of closed-loop active flow control (AFC) systems is highly complex.
As an example, consider the flow past a truck.
At highway speeds, such a flow is fully 3D and comprises turbulent boundary layers, flow separation, reattachment, and large turbulent wakes \cite{hucho1994,choi2014}.
To implement closed-loop AFC, this complexity must be captured with a limited amount of sensors placed on the vehicle's surface.
Similarly, the actuators must be designed and placed sensibly to control the dominant flow structures.
Moreover, a suitable control law must be derived, i.e., a mapping from the sensor reading to the optimal actuation.
Finally, the nonlinear flow dynamics create a tight coupling between sensors, actuators, and control law.
Hence, the AFC system should be designed and optimized as a whole.

Recently, reinforcement learning (RL), and in particular deep RL (DRL), started to emerge as a technology in fluid mechanics with the potential to tackle the complexity of closed-loop AFC.
Viquerat et al. \cite{viquerat2022} provide an extensive account of DRL-based applications in fluid mechanics.
DRL solves control problems through trial-and-error interactions with an environment, where \textit{environment} is the DRL term for the system to be controlled.
The unique advantage of learning from simulation-based environments is the possibility of end-to-end optimization of the full control system, i.e., sensor locations, control law, and actuator positions \cite{paris2021, krogmann2023, paris2023}.
However, a critical challenge of simulation-based RL is the high computational cost and turnaround time of flow simulations.
Even for relatively simple flow control benchmark problems, state-of-the-art RL algorithms require $O(100)$ episodes to converge \cite{viquerat2022}, which is far from practical for real-world AFC applications.

To emphasize the need for sample-efficient learning, consider again the application of DRL-based closed-loop AFC to vehicles.
The current gold standard for accurate and fast simulations of vehicle aerodynamics are hybrid approaches like improved delayed detached eddy simulations (IDDES) \cite{ashton2016}.
Computing converged mean force coefficients for the \textit{DrivAer} test case at $Re=768000$, required approximately two days on 700 CPU cores in 2016 \cite{ashton2016}.
It is conceivable that roughly five sequential IDDES can be performed per day because it is not necessary to obtain converged statistics in every run.
Moreover, hardware and scalability have improved significantly over the past years.
Assuming that 10 simulations are performed in parallel per episode to generate more data, each simulation being run on 1000 CPU cores, the computational cost for 100 episodes is approximately 5M CPU hours, and the training lasts 20 days.
Cloud providers charge 0.01-0.05 euros per CPU hour, which puts a price tag of 0.5M-2.5M euros on a single training at a single flow condition.
Additional hyperparameter optimization, sensor/actuator variants, and an extension to multiple flow conditions can easily increase the cost by a factor of 10-100.
This cost, combined with a turnaround time of 20 days, is far from practical for the majority of potential users.

Several remedies exist to reduce the computational cost and turnaround time, e.g., by exploiting invariances \cite{belus2019,vignon2023} or by pre-training on simulations with very coarse meshes \cite{dixit2023}.
While effective, the applicability of the aforementioned techniques strongly depends on the control problem, i.e., the problem must exhibit invariances, and the mesh coarsening must not change the flow characteristics.
A more general approach to improve data efficiency is model-based DRL (MBDRL) \cite{moerland2022}.
The main idea is to substitute the expensive simulation-based environment with one or more surrogate models.
The surrogate models are optimized based on data coming from the high-fidelity environment.
Additional data to optimize the control law can be created with little effort by querying the optimized surrogate model.
A plethora of MBDRL algorithms exist that differ in the type of surrogate model and how the control law is derived from the model.
Moerland et al. \cite{moerland2022} provide an extensive overview.

Key challenges in MBDRL are i) the efficient creation of accurate surrogate models and ii) dealing with the presence of model error.
Due to their flexibility, neural networks are a common choice for building data-driven surrogate models.
However, the models must be created on the fly based on data that is unavailable for prior tests, so the optimization pipeline must be robust and efficient.
Moreover, the models are auto-regressive.
Hence, long-term stability is a persistent challenge.
If the prediction accuracy is not sufficiently high, querying the models repeatedly leads to vanishing or exploding solutions.
Finally, even if the model creation workflow is robust and efficient, there is always epistemic uncertainty due to the alternating iterative updates of environment models and control law.
With each update of the control law, previously unknown actuations and states will emerge, and consequently, the environment models will eventually become too inaccurate.
Estimating whether the surrogate models are still reliable or should be updated with high-fidelity data from the true environment is essential to making MBDRL robust.

In this contribution, we present a modified version of the model ensemble trust region policy optimization (METRPO) \cite{kurutach2018} algorithm and demonstrate the benefits of MBDRL for fluid-mechanical applications.
Specifically, we compare model-free (MF) and model-based (MB) learning on the flow past a circular cylinder \cite{rabault2019,tokarev2020} and the fluidic pinball configuration \cite{holm2020,krogmann2023}.
The appendix describes various implementation details and serves as a reference for the abbreviations used throughout the text.
Moreover, we provide a complimentary code repository with instructions to reproduce the numerical experiments and data analyses presented hereafter \cite{articleRepo}.
A persistent record of all scientific results, including a snapshot of the code repository, is publicly accessible \cite{articleData}.

\section{Theory}\label{sec:drl_theory}
\begin{figure}[htbp]
	\begin{center}
		\includegraphics[width=0.9\textwidth]{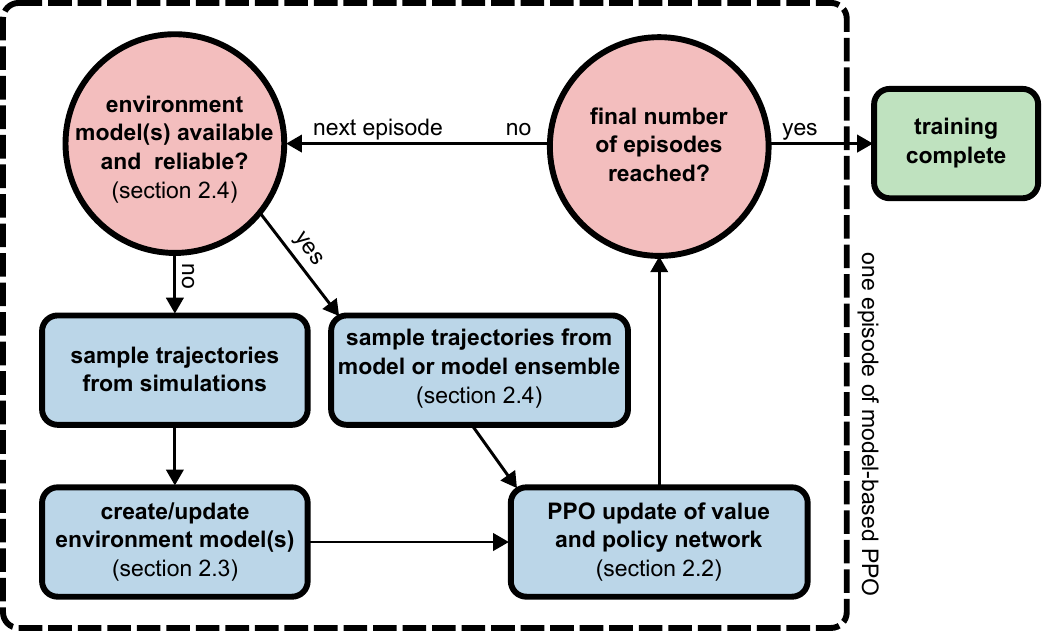}
		\caption{High-level overview of one model-based PPO episode.}
		\label{fig:mb_drl_overview}
	\end{center}
\end{figure}

This section starts with a very short introduction to essential RL concepts and then describes the MF algorithm, proximal policy optimization (PPO), employed to solve the flow control problems.
Thereafter, we explain the creation of model ensembles, which emulate the simulation outputs, and how these ensembles are employed in an MBDRL variant of PPO.
Figure \ref{fig:mb_drl_overview} shows how the different algorithmic components are connected within one episode of model-based PPO.

\subsection{Reinforcement learning}\label{sec:rl_basics}
There are two main entities in RL: the agent and the environment.
The agent contains the control logic, while the environment represents the simulation or experiment that shall be controlled.
The environment at a discrete time step $n\in \mathbb{N}$ is characterized by a state $S_n\in \mathcal{S}$, where $\mathcal{S}$ is the space of all possible states.
Formally, there is a difference between the system's state and a partial observation of the state, since the full state is often inaccessible.
Nonetheless, we follow standard practice and refer to partial observations and full states alike simply as states.
In fluid mechanics, $S_n$ could represent instantaneous pressure or velocity values at one or more sensor locations in the region of interest.
The agent can manipulate the state via an action $A_n \in \mathcal{A}$, where $\mathcal{A}$ is the space of all possible actions, upon which the environment transitions to a new state $S_{n+1}$.
Common actions (means of actuation) for flows are suction, blowing, rotation, or heating.
The control objective in RL is expressed by the reward $R_{n+1}\in \mathcal{R}$, where $\mathcal{R}$ is the space of all possible rewards, and is assessed based on the transition from $S_n$ to $S_{n+1}$.
For example, if the control objective is drag reduction, the reward would be defined such that large drag forces yield low rewards while small drag forces yield high rewards.
The combination of $S_n$, $A_n$, $R_{n+1}$, and $S_{n+1}$ is called an experience tuple \cite{sutton2018}.
The aggregation of experience tuples over $N$ interactions between agent and environment forms a so-called trajectory \cite{sutton2018}, $\tau = \left[ (S_0, A_0, R_1, S_1), \ldots ,(S_{N-1}, A_{N-1}, R_N, S_N)\right]$.
Trajectories form the basis for optimizing the policy $\pi (A_n=a|S_n=s)$ (the control law), which predicts the probability of taking some action $a$ given that the current state is $s$.
Optimality in RL is defined in terms of cumulative rewards, the so-called return $G_n = \sum_{i=n+1}^N \gamma^i R_i$, because the system's response to an action may be delayed.
The discount factor $\gamma \in (0, 1]$ introduces a notion of urgency, i.e., the same reward counts more if achieved early in a trajectory.
Given the same policy $\pi$  and the same state $S_n$, the return $G_n$ might vary between trajectories.
Incomplete knowledge of the state is one common source of uncertainty.
To consider uncertainty in the optimization, RL ultimately aims to find the policy that maximizes the expected return $v_\pi (s)$ under policy $\pi$ over all possible states $s\in\mathcal{S}$ \cite{sutton2018}:
\begin{equation}
\label{eq:opt_policy}
    v_\pi^\ast (s) = \underset{\pi}{\mathrm{max}}\ \underbrace{\mathbb{E}_\pi[G_n|S_n=s]}_{v_\pi(s)},
\end{equation}
where $\mathbb{E}_\pi[\cdot]$ denotes the expected value under policy $\pi$, e.g., the mean value for a normally distributed random variable, and $v_\pi^\ast (s)$ is the optimal state-value function.
Equation \eqref{eq:opt_policy} forms the basis for all RL algorithms.

\subsection{Policy learning}\label{sec:ppo_theory}

Expression \eqref{eq:opt_policy} implicitly defines the optimal policy $\pi^\ast (a|s)$, which is the policy associated with the optimal value function $v^\ast_\pi (s)$.
However, additional algorithmic building blocks are required to derive a workable policy.
PPO is the most common algorithm to solve problem \eqref{eq:opt_policy} for fluid-mechanical systems \cite{viquerat2022}.
PPO employs deep neural networks as ansatz for the value function $v_\pi (s)$ and the policy $\pi (a|s)$.
Hence, PPO belongs to the group of actor-critic DRL algorithms.
PPO is relatively straightforward to implement and enables accelerated learning from multiple trajectories, i.e., multiple trajectories may be sampled and processed in parallel rather than sequentially to produce new training data $T = \left[ \tau_1, \tau_2,\ldots \tau_K \right]$.
It is important to note that there are many PPO details that vary across different implementations.
Many of these details are not thoroughly described in the standard PPO reference article \cite{schulman2017}, but they are nonetheless of utmost importance.
Andrychowicz et al. \cite{andrychowicz2021} provide guidelines based on comprehensive test results for a multitude of algorithmic variants.
Our implementation is based on the textbook by Morales \cite{morales2020}.
We outline only those elements needed for the MBDRL version.

Based on the trajectories, the free parameters $\theta_v$ of the parametrized value network $v_{\theta_v} (s)$ are optimized to predict the expected return over all trajectories.
The value function's subscript $\pi$ has been dropped to simplify the notation.
The loss function includes some additional clipping to limit the change between two updates \cite{morales2020}:
\begin{align}
    \label{eq:value_loss}
    L_\mathrm{val} &= \frac{1}{KN}\sum\limits_{k=1}^K \sum\limits_{n=0}^{N-1} \mathrm{max}\left\{
    \left(G_n^k - v_{\theta_v}(S_n^k)\right)^2, 
    \left(G_n^k - V_{\mathrm{clip},n}^k\right)^2\right\},\\
    V_{\mathrm{clip},n}^k &= v^\mathrm{old}_{\theta_v}(S_n^k)+\mathrm{clip}\left\{v_{\theta_v}(S_n^k)-v^\mathrm{old}_{\theta_v}(S_n^k),-\delta , \delta \right\}, \nonumber
\end{align}
where $v^\mathrm{old}_{\theta_v}$ is the value network's state before optimization, and the clip function is defined as:
\begin{equation}
    \label{eq:clip}
    \mathrm{clip}(x, -\delta, \delta) = \begin{cases}
    x, & \text{for } -\delta \leq x \leq \delta \\
    -\delta, & \text{for } x < -\delta \\
    \delta, & \text{for } x > \delta
  \end{cases}.
\end{equation}

Similar to the value function network, the policy network $\pi_{\theta_\pi} (a|s)$ parametrizes a multivariate Beta distribution over possible actions.
During exploration, actions are sampled at random from this distribution, i.e., $A_n\sim \pi_{\theta_\pi} (S_n)$.
We do not consider a potential cross-correlation between actions for exploration.
To evaluate if a sampled action performs better than expected, the reward is compared to the expected reward, i.e. \cite{schulman2017,morales2020}:
\begin{equation}
    \label{eq:step_advantage}
    \delta_n = R_n - (v_{\theta_v} (S_n) - \gamma v_{\theta_v}(S_{n+1})),
\end{equation}
where $\delta_n$ is called the 1-step advantage estimate at time step $n$.
Actions with a positive advantage estimate should be favored.
Rather than looking one step ahead, one could also look $l$ steps ahead to consider delayed effects.
The generalized advantage estimate $\hat{A}$ balances between short-term and long-term effects using the hyperparameter $\lambda \in (0,1]$ \cite{schulman2017,morales2020}:
\begin{equation}
    \label{eq:advantage}
    \hat{A}_n = \sum\limits_{l=0}^{N-n} (\gamma\lambda)^l \delta_{n+l}.
\end{equation}
Finally, the free parameters of the policy network, $\theta_\pi$, are optimized to make favorable actions ($\hat{A}_n > 0$) more likely and poor actions ($\hat{A}_n \leq 0$) less likely. This concept is combined with a clipping mechanism that restricts policy changes \cite{morales2020}:
\begin{align}
    L_\mathrm{pol} &= -\frac{1}{KN}\sum\limits_{k=1}^K \sum\limits_{n=0}^{N-1} \mathrm{min}\left\{
    \frac{\pi_{\theta_\pi} (A_n^k|S_n^k)}{\pi_{\theta_\pi}^{\mathrm{old}} (A_n^k|S_n^k)}\hat{A}_n^k, 
    \mathrm{clip}\left\{
    \frac{\pi_{\theta_\pi} (A_n^k|S_n^k)}{\pi_{\theta_\pi}^{\mathrm{old}}(A_n^k|S_n^k)}, -\varepsilon, \varepsilon \right\}\hat{A}_n^k\right\}\nonumber\\
    &- \frac{\beta}{KN}\sum\limits_{k=1}^K \sum\limits_{n=0}^{N-1} E\left[\pi_{\theta_\pi}(S_n^k)\right]\label{eq:policy_loss}
\end{align}
where $\pi_{\theta_\pi}/\pi_{\theta_\pi}^\mathrm{old}$ is the probability ratio between the current and old (before optimization) policy, $\varepsilon$ is the clipping parameter, $E(\pi)$ is the policy's entropy \cite{morales2020}, and $\beta$ is the entropy's weighting factor.
The additional entropy term avoids a premature stop of the exploration \cite{moerland2022}.

One sequence of generating trajectories, updating the value network according to equation \eqref{eq:value_loss}, and updating the policy according to equation \eqref{eq:policy_loss} is called an episode.
Since the optimization starts with randomly initialized value and policy networks, multiple episodes are required to find the optimal policy.

\subsection{Model learning}\label{sec:model_learning}

The type of environment model employed here is a simple feed-forward neural network with weights $\theta_m$ that maps from the last $d+1$ states and a given action to the next state and the received reward:
\begin{equation}
    \label{eq:model_map}
    m_{\theta_m} : (S_{n-d}, \ldots, S_{n-1}, S_n, A_n) \rightarrow (S_{n+1}, R_{n+1}).
\end{equation}
To simplify the notation, we introduce the current extended state $\hat{S}_n$, which comprises the last $d+1$ visited states:
\begin{equation}
    \hat{S}_n = \left[ S_{n-d}, \ldots, S_{n-1}, S_n \right].
\end{equation}
Arranging the current extended state and the current action into a feature vector, $\mathbf{x}_n = [\hat{S}_n, A_n]$, and the next state and the received reward into a label vector, $\mathbf{y}_n = [S_{n+1}, R_{n+1}]$, the model weights $\theta_m$ can be optimized employing a mean squared error (MSE) loss of the form:
\begin{equation}
    \label{eq:model_loss}
    L_m = \frac{1}{|D|}\sum\limits_{i}^{|D|} (\mathbf{y}_i - m_{\theta_m}(\mathbf{x}_i))^2,
\end{equation}
where $D$ is a set of feature-label-pairs constructed from high-fidelity trajectories, and $|D|$ is the overall number of feature-label-pairs.
A single trajectory of length $N$ yields $|D| = N-(d+1)$ such pairs.
To obtain the model weights based on equation \eqref{eq:model_loss}, we employ state-of-the-art deep learning techniques, e.g., normalization of features and labels, layer normalization, batch training, learning rate schedule, and early stopping.
More details about the model training are provided in appendix \ref{sec:model_details}.

\subsection{Model-based proximal policy optimization}\label{sec:mbppo}

Given a trained model of the form \eqref{eq:model_map}, it becomes straightforward to sample new (\textit{fictitious}) trajectories from the model by employing the model recursively.
This process is described in detail in algorithm \ref{alg:simple_trajectory}.
The initial extended state $\hat{S}_0$ is selected from the existing high-fidelity trajectories.
Note that the action sampling makes the trajectory sampling stochastic, so even when starting from the same $\hat{S}_0$, executing algorithm \ref{alg:simple_trajectory} $K$ times generates $K$ different trajectories.
For efficiency, the sampling of multiple trajectories should be performed in parallel.

\begin{algorithm}
	\caption{Sampling of a trajectory from an environment model.}\label{alg:simple_trajectory}
	\begin{algorithmic}
		\State \textbf{Input:} $\hat{S}_0$, $\pi_{\theta_\pi}$, $m_{\theta_m}$ \Comment{initial history, policy, env. model}
        \State \textbf{Result:} $\tau$ \Comment{sampled model trajectory}
        \State $\tau \leftarrow \left[\right]$ \Comment{initialize empty trajectory}
		\State $\hat{S}_n \leftarrow \hat{S}_0$ \Comment{initialize current extended state}
		\While{$i < N$}
            \State $A_n \sim \pi_{\theta_\pi}(S_n)$ \Comment{sample an action from the current policy}
            \State $\left[S_{n+1}, R_{n+1}\right] \leftarrow m_{\theta_m}(\hat{S}_n, A_n)$ \Comment{predict the next state and the reward}
            \State $\tau \leftarrow \tau \cup \left[[ S_n, A_n, R_{n+1}, S_{n+1} \right]]$ \Comment{append experience tuple to trajectory}
            \State $\hat{S}_n\leftarrow \left[S_{n-d+1}, \ldots, S_n, S_{n+1}\right]$ \Comment{overwrite current extended state}
		\EndWhile
	\end{algorithmic}
\end{algorithm}

In principle, MBDRL can work by sampling trajectories from a single model.
However, the policy changes with each episode, which leads to newly explored states and actions.
Naturally, the environment model's prediction error increases with each episode, and it becomes increasingly challenging to sample meaningful trajectories from the model.
Therefore, it is vital to monitor the model's confidence in a prediction to decide at which point new high-fidelity episodes should be generated.
There are two pathways to include model uncertainty: one could train a fully probabilistic environment model, i.e., a model that predicts a distribution over the next possible states, or one could train an ensemble of regular models \cite{moerland2022}.
Here, we follow the METRPO algorithm \cite{kurutach2018} and train an ensemble of $N_\mathrm{m}$ simple environment models.
Each member in the ensemble $M = \left[ m_1, m_2, \ldots, m_{N_\mathrm{m}}\right]$ has a different set of weights $\theta_{i}$, which we drop from the index to simplify the notation.
The ensemble approach introduces a new hyperparameter, namely the number of models, but training the ensemble is typically easier than training a fully probabilistic network.
Moreover, it is straightforward to optimize the models in parallel.
Each model is trained as described before, i.e., employing loss function \eqref{eq:model_loss}.
Even though the training procedure stays the same, the optimized models differ from one another for several reasons:
\begin{enumerate}
    \item The dataset $D$ is split randomly into $N_\mathrm{m}$ subsets, and each model is trained on a different subset.
    \item Each model has a different set of randomly initialized weights. Since the optimization is nonlinear, the optimized weights likely differ, even if the same dataset is used for training.
    \item The gradient descent algorithm updates the model weights multiple times per epoch (iteration) based on batch gradients. The mini-batches are chosen at random from the training data (batch training).
\end{enumerate}

There are several options to generate new trajectories from the model ensemble.
The most obvious one is to repeat algorithm \ref{alg:simple_trajectory} once per model.
In contrast, the original METRPO algorithm mixes the models within the trajectory sampling to improve robustness \cite{kurutach2018}.
The authors state that different models are likely to have different model biases due to the training on a different subset of the data.
When mixing the models, the different biases can partially cancel out.
The alternation between models is achieved by sampling a different model for each step from a categorical distribution $P_M$ over $N_\mathrm{m}$ classes, where each class $i$ has an equal selection probability of $p_i = 1/N_\mathrm{m}$.
The modified sampling strategy, which we also employ here, is depicted in algorithm \ref{alg:ensemble_trajectory}.

\begin{algorithm}
	\caption{Sampling of a trajectory from an ensemble of environment models. Differences compared to algorithm \ref{alg:simple_trajectory} are marked in bold.}\label{alg:ensemble_trajectory}
	\begin{algorithmic}
		\State \textbf{Input:} $\hat{S}_0$, $\pi_{\theta_\pi}$, $M$ \Comment{initial history, policy, \textbf{model ensemble}}
        \State \textbf{Result:} $\tau$ \Comment{sampled model trajectory}
        \State $\tau \leftarrow \left[\right]$ \Comment{initialize empty trajectory}
		\State $\hat{S}_n \leftarrow \hat{S}_0$ \Comment{initialize current extended state}
		\While{$i < N$}
            \State $A_n \sim \pi_{\theta_\pi}(S_n)$ \Comment{sample an action from the current policy}
            \State $m_n \sim P_M$ \Comment{\textbf{randomly select a model from the ensemble}}
            \State $\left[S_{n+1}, R_{n+1}\right] \leftarrow m_n(\hat{S}_n, A_n)$ \Comment{predict the next state and the reward}
            \State $\tau \leftarrow \tau \cup \left[[ S_n, A_n, R_{n+1}, S_{n+1} \right]]$ \Comment{append experience tuple to trajectory}
            \State $\hat{S}_n\leftarrow \left[S_{n-d+1}, \ldots, S_n, S_{n+1}\right]$ \Comment{overwrite current extended state}
		\EndWhile
	\end{algorithmic}
\end{algorithm}

Besides the robust generation of trajectories, the ensemble allows to quantify prediction uncertainty.
More precisely, we would like to determine, at which point the ensemble becomes unsuitable to generate new trajectories.
Therefore, in each model-based episode, $K$ trajectories are generated from the ensemble employing algorithm \ref{alg:ensemble_trajectory}, and $N_\mathrm{m}$ additional trajectories, one for each model, are generated employing algorithm \ref{alg:simple_trajectory}.
The ensemble trajectories are used to update policy and value networks.
The model-specific trajectories are used to evaluate the policy loss \eqref{eq:policy_loss} individually for each model to obtain a scalar value expressing the model's quality (not for gradient descent).
Comparing the $i$th model's loss values of the current episode, i.g., $L_\mathrm{pol,i}^\mathrm{new}$, with the loss of the same model obtained in the previous episode, i.e., $L_\mathrm{pol,i}^\mathrm{old}$, allows assessing the quality of the policy update in the current episode.
The loss trends for all models in the ensemble are combined by evaluating:
\begin{equation}
    \label{eq:confidence}
    N_\mathrm{pos} = \sum\limits_{i=1}^{N_\mathrm{m}} H\left( L_\mathrm{pol,i}^\mathrm{old} - L_\mathrm{pol,i}^\mathrm{new} \right),
\end{equation}
where $H$ is the Heaviside step function.
In simple terms, $N_\mathrm{pos}$ is the number of models in the ensemble for which the policy loss improves after an update of the policy.
Values of $N_\mathrm{pos}/N_\mathrm{m}$ close to zero indicate that the models' generalization capabilities are exhausted.
Consequently, new high-fidelity data should be generated to update the ensemble.
We define the condition $N_\mathrm{pos} \ge N_\mathrm{thr}$ to switch between model-based and simulation-based trajectory sampling.
Values of $N_\mathrm{thr}/N_\mathrm{m}$ close to zero encourage learning from the models at the risk of decreased robustness, whereas values close to one ensure high model quality at the risk of decreased computational efficiency.
It is conceivable that $N_\mathrm{thr}/N_\mathrm{m} \approx 0.5$ might be a suitable compromise between efficiency and control performance.
The final model ensemble PPO (MEPPO) workflow is depicted in algorithm \ref{alg:meppo}.

\begin{algorithm}
	\caption{Model-based proximal policy optimization (MEPPO) algorithm.}\label{alg:meppo}
	\begin{algorithmic}
        \State \textbf{Input:} $v_{\theta_v}$, $\pi_{\theta_\pi}$, $M$ \Comment{initial value and policy networks, model ensemble}
        \State \textbf{Result:} $\theta_\pi^\ast$ \Comment{optimal policy weights}
        \State $e \leftarrow 0$ \Comment{initialize episode counter}
		\While{$e < e_\mathrm{max}$}
		\If{$N_\mathrm{pos} < N_\mathrm{thr}$ \textbf{or} $e < 2$}								\Comment{sample high-fidelity data}
		\State $T \leftarrow \mathrm{GenerateCFDTrajectories}(\pi_{\theta_\pi})$ \Comment{run $K$ simulations}
		\State $M\leftarrow \mathrm{UpdateModelEnsemble}(M, T)$ \Comment{loss function \eqref{eq:model_loss}}
        \State $N_\mathrm{pos} \leftarrow N_\mathrm{m}$ \Comment{all models trustworthy}
		\Else \Comment{sample from model ensemble}
		\State $T \leftarrow \mathrm{GenerateEnsembleTrajectories}(\pi_{\theta_\pi}, M)$ \Comment{use algorithm \ref{alg:ensemble_trajectory} $K$ times}
		\State $N_\mathrm{pos} \leftarrow \mathrm{EvaluateModels}(\pi_{\theta_\pi}, M)$ \Comment{algorithm \ref{alg:simple_trajectory}, equation \eqref{eq:confidence}}
		\EndIf
        \State $\pi_{\theta_\pi} \leftarrow \mathrm{UpdatePolicyNet}(\pi_{\theta_\pi}, v_\mathrm{\theta_v}, T)$ \Comment{loss function \eqref{eq:policy_loss}}
		\State $v_\mathrm{\theta_v} \leftarrow \mathrm{UpdateValueNet}(v_\mathrm{\theta_v}, T)$ \Comment{loss function \eqref{eq:value_loss}}
		\EndWhile
	\end{algorithmic}
\end{algorithm}

\newpage
\section{Results}\label{sec:results}
We demonstrate the MEPPO algorithm on two flow configurations, namely a rotating cylinder in channel flow and the fluidic pinball.
The numerical setups are described in section \ref{sec:results:subsec:setup}.
In section \ref{sec:results:subsec:results}, we compare MF and MB trainings and investigate the influence of ensemble size $N_\mathrm{m}$ and switching criterion $N_\mathrm{thr}$.
Section \ref{sec:results:subsec:analysis} presents a short discussion of the optimal policies found for each test case.
The numerical simulations are performed with the \textit{OpenFOAM-v2206} toolbox \cite{OpenFOAMv2206}.
The orchestration of simulations and the DRL logic are implemented in the \textit{drlFoam} package \cite{drlFoam}.
Further implementation details are available in the complementary code repository \cite{articleRepo}.


\subsection{Flow control problems}\label{sec:results:subsec:setup}
\subsubsection{Cylinder flow}\label{sec:setup_cylinder}

The flow past a circular cylinder has become an established AFC benchmark.
Originally, the setup was introduced by Sch\"{a}fer et al. \cite{schaefer1996} as a benchmark for numerical methods.
Rabault et al. \cite{rabault2019} were the first to adopt the setup for DRL-based AFC.
Since then, many variants with different means of actuation, sensor positions, and Reynolds numbers have been investigated \cite{viquerat2022}.

\begin{figure}[htbp]
	\begin{center}
		\includegraphics[width=0.9\textwidth]{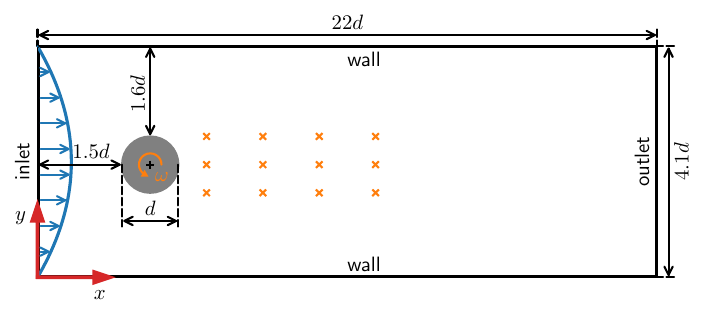}
		\caption{AFC setup of the flow past a cylinder; based on \cite{schaefer1996,rabault2019,tokarev2020}.}
		\label{fig:setup_cylinder}
	\end{center}
\end{figure}

The setup used in the present study is depicted in figure \ref{fig:setup_cylinder}.
The inlet velocity profile is parabolic \cite{schaefer1996}, while the velocity vector at the upper and lower domain boundary is zero.
At the outlet boundary, the velocity gradient is set to zero.
For the pressure, a fixed reference value is applied at the outlet, while the gradient is set to zero on all other boundaries.
The Reynolds number based on the mean inlet velocity $U_\mathrm{in}$, the cylinder diameter $d$ and the kinematic viscosity $\nu$ is $Re=U_\mathrm{in}d/\nu = 100$.
To solve the incompressible Navier Stokes equations, we employ the \textit{pimpleFoam} solver with residual control.
The pressure-velocity coupling is stopped once the initial residuals for pressure and momentum equations drop below $10^{-4}$.
Since the flow is laminar, no additional turbulence modeling is necessary.
The mesh consists of approximately $5.3\times 10^3$ hexahedral cells and is created using the \textit{blockMesh} utility.
For completeness, we note that the mesh has one cell layer in the third spatial direction of depth $\Delta z = 0.1d$, even though the simulation is 2D.
The extension in the third direction is a technical requirement of 2D simulations in \textit{OpenFOAM}.
To discretize convective and diffusive fluxes, we employ pure linear interpolation.
An implicit first-order Euler method is used for the temporal discretization.

The state (observation) used as input for control is formed by 12 pressure probes placed in the cylinder's wake.
The flow is actuated by rotating the cylinder.
The control aims to minimize the forces acting on the cylinder.
Given the density-normalized integral force vector $\mathbf{F} = \left[F_x, F_y\right]^T$, the corresponding force coefficients along each coordinate are:
\begin{equation}
    c_x = \frac{2F_x}{U_\mathrm{in}^2 A_\mathrm{ref}},\quad c_y = \frac{2F_y}{U_\mathrm{in}^2 A_\mathrm{ref}},
\end{equation}
where the reference area is $A_\mathrm{ref} = d\Delta z$.
The instantaneous reward at time step $n$ is computed as \cite{tokarev2020}:
\begin{equation}
\label{eq:reward_cylinder}
    R_n = 3 - \left(c_{x,n} + 0.1|c_{y,n}|\right).
\end{equation}
Note that we do not use time or window-averaged coefficients but the instantaneous values $c_{i,n}$.
The \textit{magical constants} in equation \eqref{eq:reward_cylinder} simply scale the reward to a value range near zero, which avoids additional tuning of PPO hyperparameters.

Given the convective time scale $t_\mathrm{conv} = d/U_\mathrm{in}$, the control starts once the quasi-steady state is reached, which is at $40t_\mathrm{conv}$.
The policy is queried once every 20 numerical time steps, which corresponds to a control time step of $\Delta t_\mathrm{c} = 0.1t_\mathrm{conv}$.
Within the control interval, the angular velocity linearly transitions from $\omega_n$ to $\omega_{n+1}$ \cite{tokarev2020}.
The value range of the normalized angular velocity, $\omega^\ast = \omega d/U_\mathrm{in}$, is limited to $\omega^\ast \in \left[-0.5, 0.5\right]$.
Overall, $N=400$ experience tuples are generated within one trajectory, which corresponds to a duration of $40t_\mathrm{conv}$.
The generation of a single trajectory takes approximately $4min$ on two MPI ranks.

\subsubsection{Fluidic pinball}
\label{sec:setup_pinball}

Noack et al. \cite{noack2016} introduced the fluidic pinball, which is a triangular configuration of three rotating cylinders.
The unforced flow undergoes several different vortex shedding regimes, namely symmetric, asymmetric, and chaotic vortex shedding.
The cylinder-based Reynolds number of the present setup is $Re=100$, which places the flow in the asymmetric vortex shedding regime.

\begin{figure}[ht]
	\begin{center}
		\includegraphics[width=0.9\textwidth]{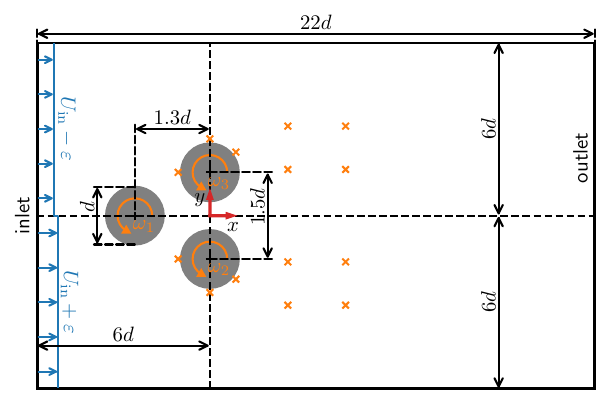}
		\caption{AFC setup of the fluidic pinball; based on \cite{noack2016}.}
		\label{fig:setup_pinball}
	\end{center}
\end{figure}

The setup is depicted in figure \ref{fig:setup_pinball}.
To reduce the initial transient simulation phase, we apply a step function velocity profile at the inlet with $\varepsilon = 0.01U_\mathrm{in}$.
A constant velocity vector is set on the lower and upper domain boundaries, i.e., $1.01U_\mathrm{in}$ and $0.99 U_\mathrm{in}$, respectively.
The velocity gradient at the outlet boundary is set to zero.
Similar to the cylinder setup, the pressure is fixed at the outlet, and the pressure gradient on all other boundaries is set to zero.
The remaining details of the setup are largely the same as those presented in section \ref{sec:setup_cylinder}.
Due to the more complex geometry, the mesh consists of approximately $5.3\times 10^4$ control volumes.
For the convective term of the momentum equation, we employ a \textit{limitedLinear} scheme with a coefficient of $1.0$.

A total of 14 pressure sensors form the state.
The positions depicted in figure \ref{fig:setup_pinball} are loosely inspired by a preliminary optimal sensor placement study \cite{krogmann2023}.
The precise numerical coordinates of each sensor can be inferred from the setup files in the complementary code repository.
The policy parametrizes a trivariate Beta distribution over the angular velocities $\omega_i$, $i\in \lbrace 1, 2, 3 \rbrace$.
As the control objective, we aim to minimize the cumulative forces acting on all three cylinders.
Given the density-normalized integral force $\mathbf{F}_i = \left[F_{x,i}, F_{y,i}\right]^T$ of the $i$th cylinder, the corresponding force coefficients are defined as:
\begin{equation}
\label{eq:pinball_coeff_i}
    c_{x,i} = \frac{2F_{x,i}}{U_\mathrm{in}^2 A_\mathrm{ref}},\quad c_{y,i} = \frac{2F_{y,i}}{U_\mathrm{in}^2 A_\mathrm{ref}},
\end{equation}
where the reference area is the projected area of all three cylinders in $x$-direction, namely $A_\mathrm{ref} = 2.5d\Delta z$.
Definition \eqref{eq:pinball_coeff_i} ensures that the sum over all cylinders recovers the correct total force coefficients:
\begin{equation}
    \label{eq:pinball_coeff}
    c_x = \sum\limits_{i=1}^3 c_{x,i},\quad c_y = \sum\limits_{i=1}^3 c_{y,i}.
\end{equation}
Based on the cumulative force coefficients, the instantaneous reward is computed as:
\begin{equation}
\label{eq:reward_pinball}
    R_n = 1.5 - (c_{x,n} + 0.5 |c_{y,n}|).
\end{equation}
The reward definition is very similar to the one used for the single cylinder.
Note that there is no particular reason for this definition other than that it was employed in a previous study \cite{holm2020}.
Alternatively, we could have also summed up the magnitudes of the individual force coefficients.
However, the focus of this work is on the demonstration of MBDRL.
The \textit{magical constants} in equation \ref{eq:reward_pinball} have the same purpose as before, namely normalizing the reward values.

Keeping the same definition of the convective time scale as before, the control starts at the end of the initial transient regime, which is at $200t_\mathrm{conv}$.
The policy is queried every 50 numerical time steps, corresponding to a control interval of $\Delta t_c = 0.5t_\mathrm{conv}$.
The trajectory extends over $100t_\mathrm{conv}$ such that $N=200$ experience tuples are generated.
The dimensionless angular velocities are limited to the range $\omega^\ast_i \in \left[-5, 5\right]$.
Generating a single trajectory with eight MPI ranks requires approximately $40min$.
The main parameters relevant to AFC are summarized in table \ref{table:setup} for both test cases.

\begin{table}[ht]
	\caption{Characteristic parameters of the two AFC setups; \textit{ranks} refers to the number of MPI ranks; \textit{CV} is the number of control volumes, and $T_\mathrm{tr}$ is the time required to execute one simulation (to sample one trajectory).}\label{table:setup}%
	\begin{tabular}{@{}lcccccccc@{}}
		\toprule
		case & $Re$ & $t_\mathrm{conv}$ & $\Delta t_c/t_\mathrm{conv}$ & $\omega^\ast$ range & $N$ & ranks & CV & $T_\mathrm{tr}$ \\
		\midrule
		cylinder & $100$ & $0.1s$ & $0.1$ & $\left[-0.5, 0.5\right]$ & $400$ & $2$ & $5.3\times 10^3$ & $\approx 4min$ \\
		pinball & $100$ & $1s$ & $0.5$ & $\left[-5, 5\right]$ & $200$ & $8$ & $5.3\times 10^4$ & $\approx 40min$ \\
		\botrule
	\end{tabular}
\end{table}

\subsection{Training performance}\label{sec:results:subsec:results}

\subsubsection{General remarks about the training}
\label{sec:general_training_remarks}
We compare the control performance and the computational cost between MF and MB trainings.
Each training is executed in isolation on a dedicated compute node with 96 CPU cores.
To discuss the schematics presented later on, it is necessary to outline several implementation details.
In each episode, exactly ten trajectories are generated in parallel before updating the policy and value networks.
The \textit{buffer size} and the parallel trajectory generation are the same for CFD and MB trajectories.
We perform a very rough check of the force coefficients in the MB trajectories and discard trajectories that are obviously extremely inaccurate.
Specifically, we only keep an MB trajectory for the cylinder flow if all coefficients fulfill the criteria $2.85 \le c_x \le 3.5$ and $|c_y| \leq 1.3$.
The analogous bounds for pinball trajectories are $-3.5 \le c_{x,i} \le 3.5$ and $|c_{y,i}| \leq 3.5$.
Note that these bounds are really generous and filter out only extreme prediction errors.
If one or more trajectories are discarded, the sampling is repeated until ten valid trajectories are available.
The training is stopped after 200 and 150 episodes for the cylinder and pinball cases, respectively.
Finally, we note that all parameter configurations for the cylinder flow are repeated with five different seed values.
All results presented in section \ref{sec:perf_cylinder} are averaged over these five runs.
Due to the computational cost of the fluidic pinball, we execute only a single training run per parameter configuration.

\subsubsection{Cylinder flow}

\begin{figure}[htbp]
	\begin{center}
		\includegraphics[width=\textwidth]{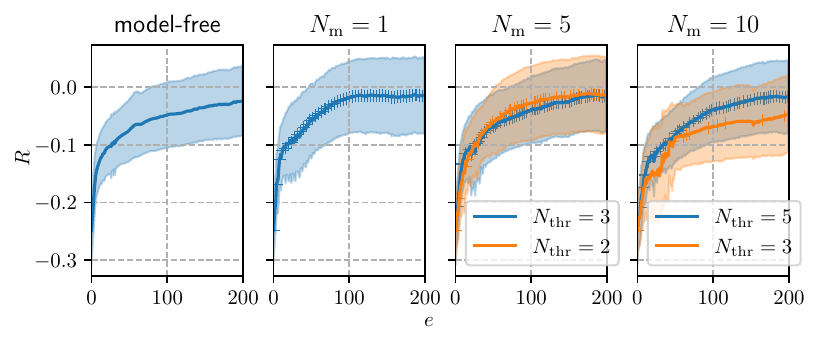}
		\caption{Cylinder flow: episode-wise mean reward $R$ for different training configurations; the shaded area encloses one standard deviation below and above the mean; mean and standard deviation are computed over all trajectories of all seeds; for the MB training, the markers indicate CFD-based trajectory sampling.}
		\label{fig:avg_rewards_cylinder}
	\end{center}
\end{figure}

Figure \ref{fig:avg_rewards_cylinder} shows the episode-wise mean rewards received with different training configurations.
Within 200 episodes almost all configurations achieve a similar maximum reward.
The maximum reward in the MF training is slightly lower, but the reward still keeps increasing moderately toward the end of the training.
Surprisingly, all MB trainings reach the maximum reward earlier than the MF training.
We noticed that the MB trajectories display less small-scale variance than the CFD trajectories.
Presumably, this indirect \textit{filtering} performed by the environment models affects the policy and value network updates positively.
In general, The MB trainings display a higher variance in the mean reward, especially within the first 50 episodes, where the changes in the reward are the strongest.
This behavior is related to individual MB trajectories with poor prediction quality.
Luckily, these trajectories seem to have a limited impact on the learning.

\label{sec:perf_cylinder}

The markers in the MB training in figure \ref{fig:avg_rewards_cylinder} indicate CFD trajectories.
For the MB training with one model, we switch back to simulation-based sampling every fourth episode, so $75\%$ of all episodes use the environment model.
The same ratio results in the MEPPO training with $N_\mathrm{m}=5$ and $N_\mathrm{thr}=3$.
Interestingly, the automated switching between CFD and model sampling leads to a relatively regular alternation.
The same number of models with a lowered threshold value of $N_\mathrm{thr}=2$ reduces the amount of CFD episodes to $15\%$, notably without performance degradation.
The training with $N_\mathrm{m}=10$ and $N_\mathrm{thr}=5$ also employs the model ensemble in $75\%$ of the episodes.
Reducing the threshold value in the latter case to $N_\mathrm{thr}=3$ reduces the number of CFD episodes to $6\%$.
However, the optimization becomes significantly less stable and does not reach optimal control performance.
Note that the amount of training data per model decreases as the number of models increases.
Therefore, increasing the number of sampled trajectories per episode or allowing overlapping training data could potentially stabilize trainings with a lowered threshold value.
The increased amount of training data when employing only a single model could explain the quick convergence of the corresponding training after approximately 100 episodes.

\begin{figure}[htbp]
	\begin{center}
		\includegraphics[width=\textwidth]{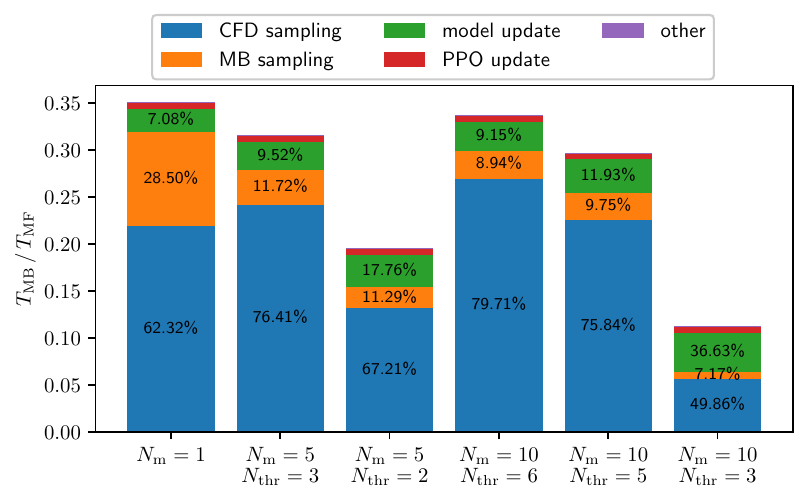}
		\caption{Cylinder flow: composition of the total MB training time $T_\mathrm{MB}$ normalized with the MF training time $T_\mathrm{MF}\approx 14h$.}
		\label{fig:exec_times_cylinder}
	\end{center}
\end{figure}

Assuming that the creation of environment models is reliable and much less costly than a CFD simulation, the more CFD episodes can be replaced with MB episodes, the lower the overall training time.
This trend is reflected by the normalized MB training time depicted in figure \ref{fig:exec_times_cylinder}.
Remarkably, all configurations reduce the training time by at least $65\%$, even though the computational cost to simulate the cylinder flow is fairly small.
To better understand differences in the overall training time, figure \ref{fig:exec_times_cylinder} also shows the time spent on the most relevant parts of the training.
For all configurations, the simulation runs remain the most time-consuming element.
The time spent updating the networks for environment, policy, and value function remains fairly constant.

\begin{figure}[htbp]
	\begin{center}
		\includegraphics[width=\textwidth]{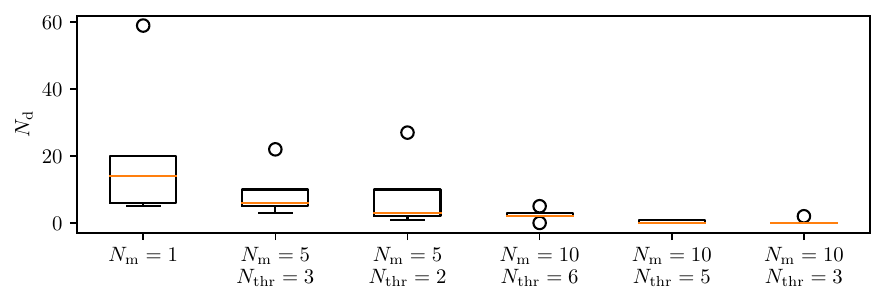}
		\caption{Cylinder flow: number of discarded trajectories $N_\mathrm{d}$ per training; the box plot shows the outcome of five independent runs (seed values); the orange line indicates the median value, and the markers indicate extreme outcomes.}
		\label{fig:discards_cylinder}
	\end{center}
\end{figure}

A noticeable increase can be observed in the time spent to sample trajectories from a single environment model.
This increase, as well as other variations in the sampling time, are related to the removal of obviously corrupted trajectories, as explained in section \ref{sec:general_training_remarks}.
Figure \ref{fig:discards_cylinder} shows the number of discarded trajectories per training.
Clearly, the ensemble models lead to a significantly more robust trajectory sampling.
Moreover, ensembles with more models produce fewer invalid samples.
It is noteworthy to mention that the majority of invalid trajectories result in the initial training phase, $e< 50$, presumably because of the strong exploration and policy changes.
In our implementation, trajectories are sampled in parallel.
However, the constraint of executing the PPO update with exactly ten valid trajectories leads to repeated sampling in the case of discarded trajectories.
Of course, this extra time could be avoided by sampling more trajectories than required in parallel or by loosening the constraint.

We note that there are also small variations in the measured execution times that are not straightforward to explain.
One means to reduce these variations would be executing additional training runs with varying seeds.
However, the main trends in terms of control performance and training time are clearly visible and explainable.
Moreover, there are nontrivial coupling effects between simulations, environment models, and PPO updates.
As discussed before, policy updates differ between MF and MB trainings and also depend to some degree on ensemble size and switching criterion.
The policy employed in a simulation impacts the execution time, e.g., by subsequent changes in the pressure-velocity-coupling iterations, or the linear solver iterations.

\subsubsection{Fluidic pinball}
\label{sec:perf_pinball}

\begin{figure}[htbp]
	\begin{center}
		\includegraphics[width=\textwidth]{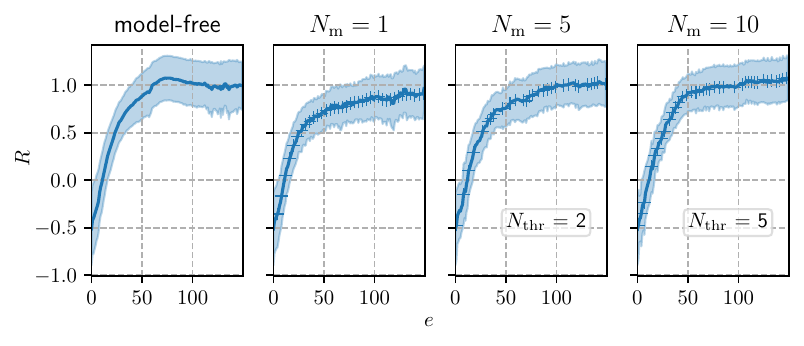}
		\caption{Fluidic pinball: episode-wise mean reward $R$ for different training configurations; the shaded area encloses one standard deviation below and above the mean; mean and standard deviation are computed over all trajectories; for the MB training, the markers indicate CFD-based trajectory sampling.}
		\label{fig:avg_rewards_pinball}
	\end{center}
\end{figure}

Due to the computational cost of the pinball setup, we do not perform multiple training runs and investigate fewer hyperparameter configurations.
Figure \ref{fig:avg_rewards_pinball} shows the episode-wise mean rewards of all trainings.
The final reward is similar for all tested configurations.
The MF training reaches the maximum reward after approximately 60 episodes.
Thereafter, the control performance drops slightly.
The highest overall reward is achieved by the MB training with $N_\mathrm{m}=10$.
For the MB training with a single model, we switch every fourth episode to CFD trajectories.
Also the automated switching criterion in the training with $N_\mathrm{m}=10$ and $N_\mathrm{thr}=5$ leads to $25\%$ CFD episodes.
The smaller ensemble with five models switches approximately every fifth episode back to the high-fidelity samples.

\begin{figure}[htbp]
	\begin{center}
		\includegraphics[width=\textwidth]{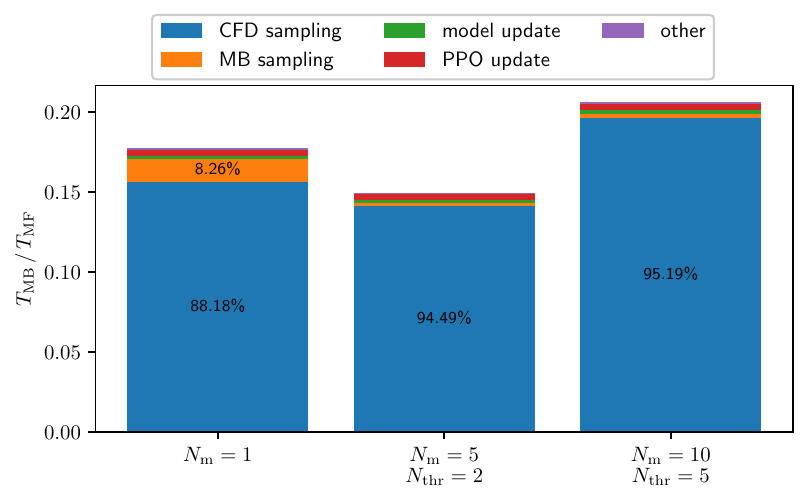}
		\caption{Fluidic pinball: composition of the total MB training time $T_\mathrm{MB}$ normalized with the MF training time $T_\mathrm{MF}\approx 130h$.}
		\label{fig:exec_times_pinball}
	\end{center}
\end{figure}

Figure \ref{fig:exec_times_pinball} shows the main contributions to the normalized training time.
Even though the relative amount of model trajectories is similar to the cylinder flow, the runtime reduction is even more significant due to the higher computational cost of performing the pinball simulation.
Remarkably, all configurations reduce the training time by at least $80\%$.
The remaining CFD episodes dominate the overall time consumption.
As for the cylinder test case, the time spent on sampling model trajectories increases when using only a single model.
Ten trajectories were discarded during the training and had to be re-sampled.
In the MEPPO runs, no invalid trajectories occurred.
The increased time spent on simulations in the training with ten models is a result of the non-trivial interaction between policy learning and the numerical properties of the simulation, as discussed before.

\subsection{Analysis of the best policies}\label{sec:results:subsec:analysis}
\subsubsection{Cylinder flow}

In this section, we compare the final policies of the MF and MB trainings that achieved the highest mean reward.
Note that the achieved force reduction and the corresponding control are by no means new \cite{viquerat2022}.
However, we would like to point out a few subtle differences in the final MF and MB policies.
It is noteworthy that all MB policies achieved a similar force reduction, even though we present only results for the MEPPO training with ten models.

\begin{figure}[htbp]
	\begin{center}
		\includegraphics[width=\textwidth]{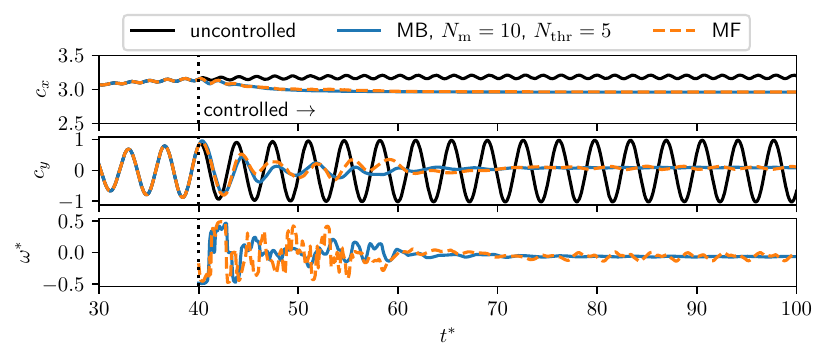}
		\caption{Cylinder flow: angular velocity and force coefficients resulting from the best MF and MB policies; the dimensionless time is $t^* = t/t_\mathrm{conv}$.}
		\label{fig:final_policies_cylinder}
	\end{center}
\end{figure}

Figure \ref{fig:final_policies_cylinder} shows the optimal control and the corresponding force coefficients.
Both MF and MB policies reduce the drag force by approximately $7\%$.
The second force coefficient remains near zero after 20 convective time units.
Both the MF and MB policies perform a very similar actuation within the first five convective time units.
Thereafter, the MB policy rotates the cylinder slightly less and archives an almost perfectly steady force balance.
Small fluctuations remain in the angular velocity and the side force when employing the MF policy.

\begin{figure}[htbp]
	\begin{center}
		\includegraphics[width=\textwidth]{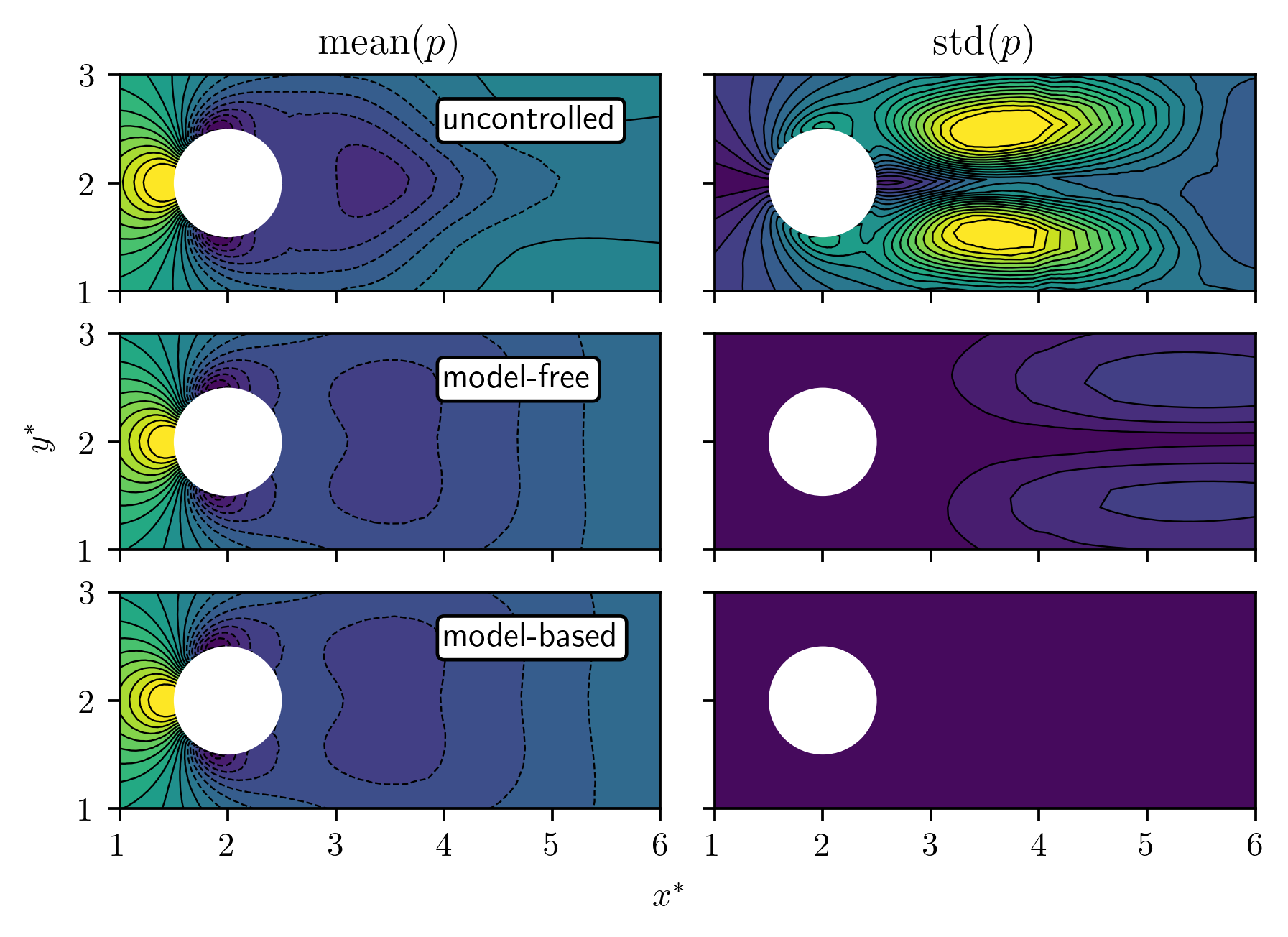}
		\caption{Temporal mean and standard deviation of the pressure fields with and without control; both mean and standard deviation are normalized with the minimum (blue) and maximum (yellow) values of the uncontrolled case; the coordinates are normalized with the diameter.}
		\label{fig:flow_fields_cylinder}
	\end{center}
\end{figure}

The cylinder rotation steadies and extends the wake.
The pressure drop over the cylinder decreases, as can be inferred visually in figure \ref{fig:flow_fields_cylinder}.
The figure also shows the significant suppression of vortex shedding.
While small pressure fluctuations remain when employing the MF policy, the MB policies leads to an almost perfectly steady flow field.
It is conceivable that the MF training could achieve a similar control performance with more training episodes and additional PPO hyperparameter tuning.
However, the MB runs achieve the presented control performance with even less than 200 episodes.

\subsubsection{Fluidic pinball}

\begin{figure}[ht]
	\begin{center}
		\includegraphics[width=\textwidth]{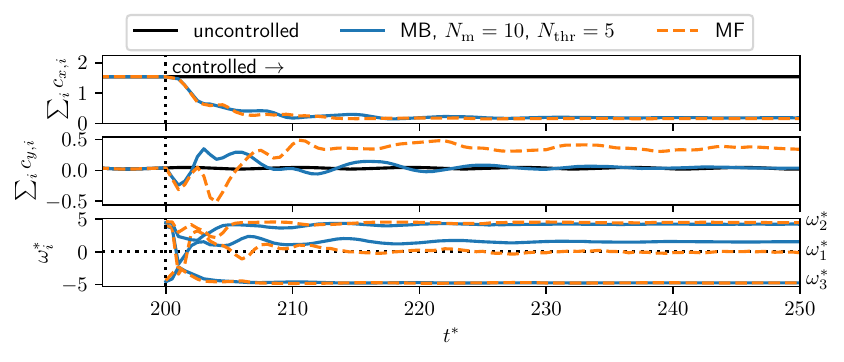}
		\caption{Fluidic pinball: angular velocities and summed force coefficients resulting from the best MF and MB policies; the dimensionless time is $t^* = t/t_\mathrm{conv}$.}
		\label{fig:final_policies_pinball}
	\end{center}
\end{figure}

Several established control mechanisms exist for the fluidic pinball, e.g., boat tailing and base bleeding \cite{raibaudo2020}.
The strategy learned by the policies in the present work is referred to as boat tailing.
Figure \ref{fig:final_policies_pinball} compares the control and the resulting force coefficients of the best-performing MF and MB policies.
Both policies follow a similar strategy, i.e., rotating cylinders 2 and 3 with maximum speed in opposite directions while keeping the cylinder in the front nearly still.
This actuation leads to boat tailing and reduces the drag coefficient by approximately $87\%$.
Of course, the reduction strongly depends on the allowed maximum rate of rotation and would be smaller for smaller rates of rotation.

While the drag reduction is fairly similar, the MF policy introduces a nonzero net force in $y$-direction.
This asymmetry is caused by a subtle difference in the absolute values of $\omega_2^\ast$ and $\omega_3^\ast$, i.e., the absolute value of $\omega_2^\ast$ is marginally larger.
The same small difference between $\omega_2^\ast$ and $\omega_3^\ast$ is present in the MB policy.
However, the MB policy applies a small positive spin to the cylinder in the front, which balances the side forces to net zero.
The reader might be wondering if the same drag reduction and net zero force in $y$-direction could be achieved by setting $\omega_1^\ast = 0$ and $\omega_2^\ast = -\omega_3^\ast = 5$.
The short answer is yes.
However, the latter control leads to relatively strong $c_y$ fluctuations whose amplitude decays only slowly.
Approximately 150 convective time units are necessary to reach the steady state with open-loop control, which is longer than the prescribed trajectory length employed in the training.
The asymmetric control performed by the MB policy achieves the same force reduction in about 20 convective time units.

\begin{figure}[htbp]
	\begin{center}
		\includegraphics[width=\textwidth]{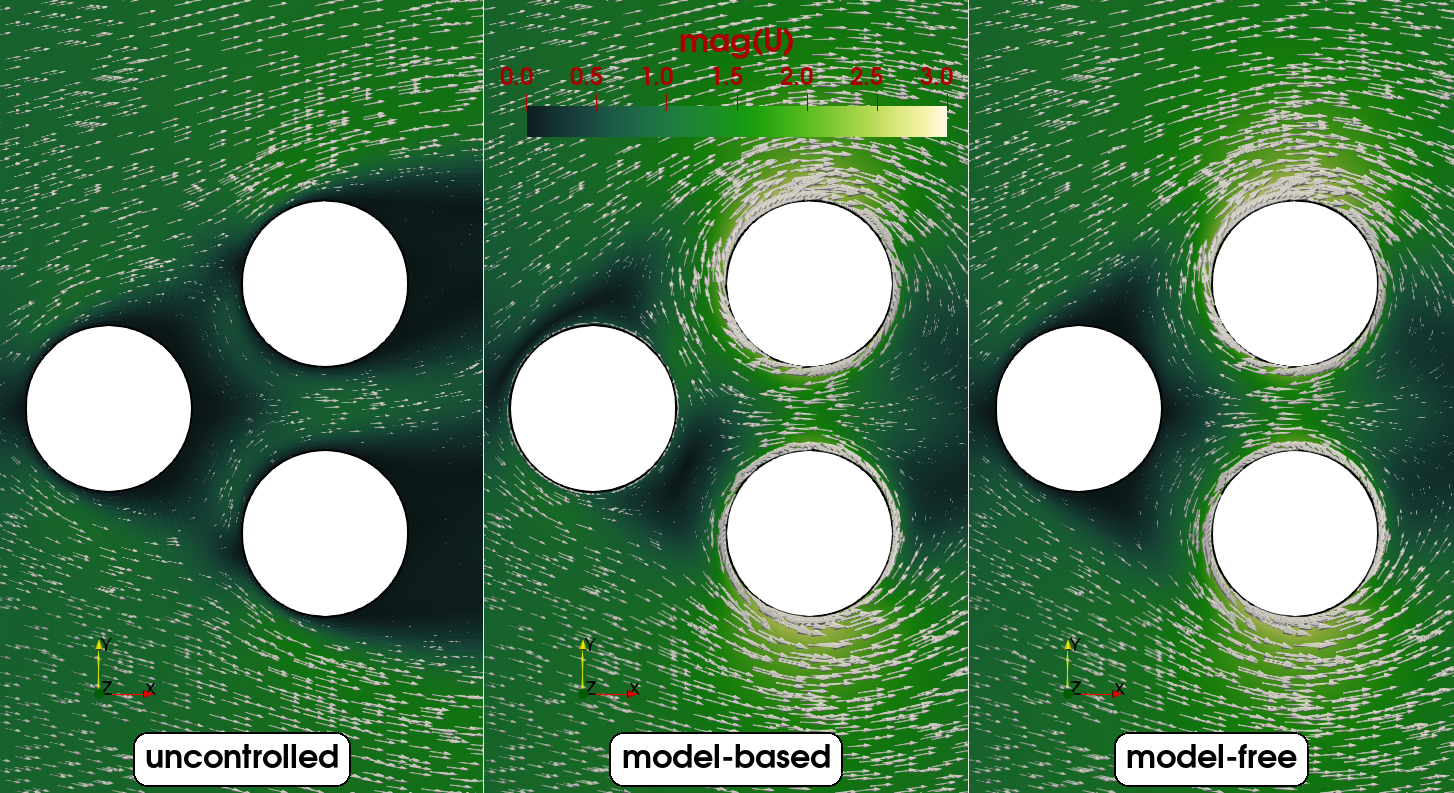}
		\caption{Comparison of instantaneous velocity fields with and without control; the velocity field is normalized with the inlet velocity.}
		\label{fig:flow_pinball}
	\end{center}
\end{figure}

Finally, we want to shed some light on the local flow dynamics created by the spinning cylinders.
Figure \ref{fig:flow_pinball} shows the velocity field in the vicinity of the cylinders.
As mentioned before, the uncontrolled state is asymmetric, as can be observed by looking at the wakes of the two cylinders in the back.
In the controlled flow, the rotation of cylinders two and three pushes fluid between them in upstream direction.
This suction significantly reduces the extent of the wake and the cumulative pressure drop over the cylinders.
The fluid then passes through the gaps between the rear cylinders and the front cylinder.
For the MF policy, the fluxes in positive and negative $y$-direction are fairly balanced.
Instead, the MB policy hinders the flux between cylinders one and two and diverts more fluid to the gap between cylinders one and three.
This small imbalance is enough to compensate for the different rotation speeds of the rear cylinders.

\section{Conclusion}\label{sec:conlusion}
DRL-based AFC is a promising approach to enable smart control technology.
A current limitation of simulation-based DRL is the high computational cost and turnaround time associated with the optimization.
The idea of MBDRL is the creation of low-cost environment models that partially replace the costly sampling of high-fidelity trajectories from simulations.
On two common benchmark AFC problems, we demonstrate that the MEPPO algorithm adopted here can tremendously reduce the training cost and time while achieving optimal control performance.
The relative reduction of training time increases with the cost of the CFD simulation.
Therefore, we expect the MEPPO algorithm, or variants thereof, to be a key enabler in performing DRL-based AFC on realistic, 3D simulations at an industrial scale.
Of course, further tests on more complex flow configurations are required to consolidate this hypothesis.

There are a number of promising options to reduce the training cost even further.
Creating accurate auto-regressive environment models on the fly posed a significant challenge.
Automating the model creation, e.g., by means of Bayesian hyperparameter optimization, could be beneficial for the models' accuracy and adaptability to new control problems.
The more accurate the models can extrapolate, the more savings are possible.
More advanced recurrent network architectures or transformers \cite{raff2022} might be capable of achieving higher accuracy than the simple, fully connected networks employed here.
However, such advanced networks are typically also harder to train.
Alternatively, more straightforward approaches to creating reduced-order models might be worthwhile to explore, too.
For example, dynamic mode decomposition with control \cite{proctor2016} could be a drop-in replacement for the models employed here.

We also noticed a strong dependency of the learning progress on the PPO hyperparameters.
Due to the high training cost, automated tuning of these parameters is infeasible for complex simulations.
However, recent improvements in the PPO algorithm, e.g., the PPO-CMA variant \cite{hamalainen2020}, reduce the number of hyperparameters, improve robustness, and accelerate learning.
Such improvements would be extremely beneficial both for MF and MB learning.

\newpage\clearpage
\backmatter

\bmhead{Supplementary information}

\bmhead{Acknowledgements}
The authors gratefully acknowledge the Deutsche Forschungsgemeinschaft DFG (German Research Foundation) for funding this work in the framework of the research unit FOR 2895 under the grant WE 6948/1-1.
The computational resources to conduct the numerical experiments were kindly provided by Amazon Web Services.
Finally, the authors gratefully acknowledge the organizing committee of the 18th OpenFOAM Workshop for the support provided during the preparation of this manuscript.

\section*{Declarations}

\subsection*{Conflict of interest}
The authors have no conflict of interest to declare.

\subsection*{Ethical approval}
No experiments on humans or animals have been conducted.

\subsection*{Funding}
This work has been funded by the German Research Foundation (DFG) under the grant WE 6948/1-1.

\subsection*{Availability of data and materials}
The full research data and instructions to reproduce the numerical experiments and analyses are available at:\\
\url{https://doi.org/10.23728/b2share.85ab8f3f68724372b83babbdaca85910}

\begin{appendices}

\section{Environment models}\label{sec:model_details}

We use a standard deep learning workflow to train the environment models.
The different optimization techniques are described in detail by Raff \cite{raff2022}.
The activation function, the number of hidden layers, and the number of neurons per layer have been determined by employing a simple grid search.
Table \ref{table:parameter_models} provides an overview of the most important hyperparameters.
Since the amount of training data is fairly small, we employ 8 CPU cores to train a single model.
Note that the models predict the individual force coefficients rather than the full reward.
Consequently, the output layer has 12+2 and 14+6 neurons for the cylinder and pinball test cases, respectively.
The training is stopped once the maximum number of epochs (gradient descent steps) is reached, the absolute loss value falls below a threshold value, or the relative change of the loss averaged over 40 epochs becomes too small.

\begin{table}[ht]
	\caption{Hyperparameters for the training of the environment models.}\label{table:parameter_models}%
	\begin{tabular}{@{}lll@{}}
		\toprule
		parameter & cylinder & pinball\\
		\midrule
		hidden layers & $3$ & $3$ \\
		layer normalization & LayerNorm & LayerNorm \\
		neurons per hidden layer & $100$ & $100$ \\
		time delays $d$ & $30$ & $30$ \\
		input neurons & $450$ & $690$ \\ 
		output neurons & $14$ & $20$ \\	
		activation function & leaky ReLU & leaky ReLU \\
		max. number of epochs & $2500$ & $2500$ \\
		batch size & $25$ & $25$ \\
		train/validation split & $75/25\%$ & $75/25\%$\\
		  absolute stopping loss & $10^{-6}$ & $10^{-6}$ \\
        relative stopping loss & $10^{-7}$ & $10^{-7}$ \\
		  loss function & MSELoss & MSELoss \\
		optimizer & AdamW & AdamW \\
		learning rate schedule & ReduceLROnPlateau & ReduceLROnPlateau \\
		  init./min. learning rate & $10^{-2}/10^{-4}$ & $10^{-2}/10^{-4}$\\
		\botrule
	\end{tabular}
\end{table}

\section{PPO hyperparameters}\label{sec:ppo_params}

Most PPO hyperparameters employed here are standard values suggested in the literature \cite{schulman2017,morales2020}.
In contrast to the original implementation, we use two separate networks for policy and value function.
The two networks are optimized sequentially by two different optimizers.
Since the optimization is inexpensive, we allocate only 5 CPU cores.
As suggested in \cite{morales2020}, the policy optimization is stopped if the difference between old and new policy, measured in terms of KL divergence, exceeds a prescribed threshold.
However, we noticed that this criterion is rarely triggered.

\begin{table}[ht]
	\caption{PPO hyperparameters.}\label{table:parameter_ppo}%
	\begin{tabular}{@{}lll@{}}
		\toprule
		parameter & cylinder & pinball \\
		\midrule
		hidden layers & $2$ & $2$ \\
		neurons per hidden layer & $64$ & $512$ \\
		input neurons & $12$ & $14$ \\
		output neurons & $2$ & $6$ \\
		activation function & ReLU & ReLU \\
		discount factor $\gamma$ & $0.99$ & $0.99$ \\
		advantage smoothing $\lambda$ & $0.97$ & $0.97$ \\
		  learning rate policy net. & $10^{-3}$ & $10^{-5}$ \\
        learning rate value net. & $5 \times 10^{-4}$ & $10^{-5}$ \\
		max. number of epochs & $100$ & $100$ \\
		batch size & full & full \\
		value and policy clipping & $0.1$ & $0.1$ \\
        optimizer & AdamW & AdamW \\
		KL divergence & $0.2$ & $0.2$ \\
		entropy weight $\beta$ & $0.01$ & $0.01$ \\
		\botrule
	\end{tabular}
\end{table}

\section{List of abbreviations}\label{sec:abbrev}

\begin{table}[]
    \centering
    \begin{tabular}{l l}
       \textbf{AFC} & active flow control \\
       \textbf{CFD} & computational fluid dynamics \\
       \textbf{DRL} & deep reinforcement learning \\
       \textbf{IDDES} & improved delayed detached eddy simulation \\
       \textbf{MB} & model-based\\
       \textbf{MEPPO} &  model ensemble proximal policy optimization \\
       \textbf{METRPO} & model ensemble trust region policy optimization \\
       \textbf{MF} & model-free\\
       \textbf{RL} & reinforcement learning \\
    \end{tabular}
    \caption{Abbreviations used throughout the article.}
    \label{tab:abbrev}
\end{table}

\end{appendices}


\bibliography{sn-bibliography}

\end{document}